\documentclass[letterpaper]{ptephy}
\usepackage{boites}
\usepackage{graphics}

\renewcommand{\titlepageheadline}{%
  \vbox to 2\baselineskip{%
    \hbox to \textwidth{%
YITP-13-79
}
\hbox to \textwidth{\rule{\textwidth}{.5pt}}
}}
\begin{document}

\title{A viable explanation of the CMB dipolar statistical anisotropy}

\author{\name{\fname{Sugumi} \surname{Kanno}}{1}, 
\name{\fname{Misao} \surname{Sasaki}}{2} 
and \name{\fname{Takahiro} \surname{Tanaka}}{2}
%\thanks{These authors contributed equally to this work}
}

\address{\affil{1}{Institute of Cosmology, Department of Physics and Astronomy, 
Tufts University, Medford, Massachusetts 02155, USA}
\affil{2}{Yukawa Institute for Theoretical Physics, Kyoto University,
Kyoto 606-8502, Japan}
\email{sugumi{}@{}cosmos.phy.tufts.edu, 
misao{}@{}yukawa.kyoto-u.ac.jp, 
tanaka{}@{}yukawa.kyoto-u.ac.jp}\\
%{\rm Report number: YITP-13-79}
}

\begin{abstract}%
The presence of a dipolar statistical anisotropy in the spectrum of
cosmic microwave background (CMB) fluctuations was reported by the Wilkinson Microwave Anisotropy Probe (WMAP), 
and has recently been confirmed in the Planck 2013 analysis of the 
temperature anisotropies. 
At the same time, the Planck 2013 results report
a stringent bound on the amplitude of the local-type non-Gaussianity.
We show that the non-linear effect of the dipolar anisotropy generates 
not only a quadrupole moment in the CMB but also a local-type non-Gaussianity.
Consequently, it is not easy to build models having a large dipolar modulation
and at the same time a sufficiently small quadrupole and level of local
bispectral anisotropy to agree with the present data.
In particular, most models proposed so far are 
almost excluded, or are at best marginally consistent with observational data.
We present a simple alternative scenario that may
explain the dipolar statistical anisotropy while satisfying 
the observational bounds on both the quadrupole moment and 
local-type non-Gaussianity. 
\end{abstract}

\subjectindex{E63, E81}

\maketitle

\section{Introduction} 
It was suggested by WMAP that the amplitude of cosmic microwave background 
(CMB) anisotropies has a dipolar directional dependence~\cite{wmap}. 
This anomalous signature has been recently confirmed by Planck~\cite{planck}. 
Although it is possible that this anisotropy is a foreground effect,
it is an interesting question whether or not we can explain this feature 
consistently within the paradigm of the inflationary universe. 

The basic issue is that a dipolar modulation of the amplitude of 
the curvature perturbation requires a seed dipolar perturbation.
This seed dipolar perturbation itself 
is known to produce quadrupole and octupole moments of CMB anisotropy,
known as the Grishchuk-Zel'dovich (GZ) effect~\cite{gz}. 
Various scenarios that explain the dipolar statistical anisotropy 
have been proposed~\cite{Kamionkowski,Liddle,Erickcek:2009at,Wang:2013lda,Liu:2013kea,McDonald:2013aca,Namjoo:2013fka,Mazumdar:2013yta,D'Amico:2013iaa,Cai:2013gma} 
and various features have been 
investigated~\cite{Lyth,Abolhasani:2013vaa,Chang:2013lxa}. 
The proposal in Ref.~\cite{Liddle} advocates the use of the 
super-curvature perturbation~\cite{Sasaki:1994yt} produced in 
the open inflation scenario %KOKO
%~\cite{GarciaBellido:1997te}
as the seed for the dipolar modulation.

The amplitudes of monopole, dipole, and 
quadrupole fluctuations $A_\ell$ $(\ell=0,1,2)$ caused by the super-curvature 
mode of a free scalar field in the small mass limit are, respectively, 
evaluated by taking the small radius limit of the corresponding normalized 
mode functions as~\cite{Sasaki:1994yt}
\begin{eqnarray}
A_0=\sqrt{3\over 2\pi}{H_{\rm F}^2\over m_{\rm F}}~, \quad 
A_1={H_{\rm F}\over 3\sqrt{\pi}} \left({r_{\rm ls}\over r_{\rm K}}\right),\quad
A_2={H_{\rm F}\over 5\sqrt{3\pi}}\left({r_{\rm ls}\over r_{\rm K}}\right)^2\,,   
\label{SCamp}
\end{eqnarray}
where $r_{\rm ls}$ and $r_{\rm K}$ are the radii of the last scattering 
surface and the curvature radius in the comoving conformal coordinate, 
and $m_{\rm F}$ and $H_{\rm F}$ are the mass and the Hubble expansion rate in the 
false vacuum before bubble nucleation. 
Thus, one can basically avoid the constraint due to the GZ
effect by choosing $r_{\rm ls}/r_{\rm K}\approx \sqrt{1-\Omega_0}$ to be
sufficiently small. 
(Although in Ref.~\cite{Lyth} a constraint on the amplitude of the
super-curvature perturbation was imposed, such a condition on the
magnitude of fluctuations beyond the current horizon scale is usually 
thought to be unnecessary in the context of the inflationary universe. 
The quasi-open inflation scenario~\cite{GarciaBellido:1997te} 
is a typical example.)

However, we suspect that one important constraint has been overlooked.
It is a genuine non-linear effect.
If the amplitude of a dipole moment is large, a non-negligible
quadrupole moment is naturally induced by non-linearity.
Taking into account the constraint from the amplitude of this 
induced quadrupole moment, 
we find that almost all models proposed so far are almost 
excluded by observations.\footnote{
A somewhat different mechanism that might circumvent this new constraint
was proposed in Ref.~\cite{D'Amico:2013iaa}.
}
In this Letter, we propose a simple alternative 
scenario that can naturally evade this constraint.

\section{Curvaton scenario}
In Refs.~\cite{Kamionkowski, Lyth}, in order to explain the 
observed large statistical anisotropy, the simplest 
curvaton scenario was studied. The only difference from the 
ordinary curvaton scenario is 
that the curvaton has a large-magnitude super-horizon fluctuation. 
In Ref.~\cite{Liddle}, this super-horizon fluctuation was 
identified with a super-curvature mode, whose amplitude can naturally be
large compared with sub-curvature modes if the creation of
the one-bubble open universe occurs in the sea of false vacuum whose
energy scale is much higher than that inside the bubble. 
As this generation mechanism of a large super-horizon fluctuation
seems most natural, we focus on possible scenarios in this context. 

First let us discuss the model based on the simplest curvaton scenario.
We denote the super-curvature fluctuation of the curvaton 
as $\Delta\chi$. 
More precisely, its distribution measured at the current 
horizon radius is $\Delta\chi \cos\theta$, where the axis of 
the angular coordinates $(\theta,\varphi)$ is aligned to 
the direction of the dipolar statistical anisotropy. %check
A typical amplitude of $\Delta\chi$ is given by $A_1$ in Eq.~(\ref{SCamp}),
whereas $\delta\chi$ represents the ordinary mode fluctuation 
whose amplitude is $O(H)$, where $H$ is the Hubble rate at the time when 
the relevant length scale crosses the horizon scale 
during the inflation. We assume $A_1\gg H$.
The inflation inside the nucleated bubble is driven by 
an inflaton field $\phi$, but the details of the model concerning 
the $\phi$-field are not important at all. 
For simplicity, we assume 
that the super-curvature perturbation is completely 
dominated by the curvaton field $\chi$ and 
that the inflaton $\phi$ is in the slow-roll regime when the relevant 
length scale crosses the horizon. 

The observed curvature perturbation can be evaluated by using 
the $\delta N$-formula~(see, for example, Ref. \cite{deltaN}), 
\begin{equation}
 {\mathcal R}_{\rm c}
 =N_\phi \delta\phi+
 N_\chi (\Delta\chi+\delta\chi)
+{1\over 2}N_{\chi\chi}(\Delta\chi+\delta\chi)^2+\cdots.
\label{deltaN}
\end{equation}
We assume that the leading-order power spectrum of the curvature
perturbation is given by the terms with the first derivatives of $N$, 
except for extremely low $\ell$ modes like dipole or quadrupole. 
Then, we have 
\begin{eqnarray*}
 P_{{\mathcal R}_c}=(N_\phi^2+N_\chi^2) \frac{H^2}{(2\pi)^2}\,. 
\end{eqnarray*}

The dipolar statistical anisotropy is caused by the term 
including $\Delta\chi$ in Eq.~(\ref{deltaN}). The leading term is given by 
\begin{eqnarray*}
 \Delta P_{{\mathcal R}_c} = 2N_\chi N_{\chi\chi} \Delta\chi 
\frac{H^2}{(2\pi)^2}\,.  
\end{eqnarray*}
The ratio of these two quantities is constrained 
by the observation as~\cite{planck}\footnote{This result, however, 
is of modest statistical significance, and most of the signal appears 
concentrated in the modulation of modes of small $\ell$. 
We thank M. Bucher for making this point clear to us.
See Ref.~\cite{planck} for a more detailed discussion.}
\begin{equation}
{\Delta P_{\delta T}\over P_{\delta T}}= {2N_{\chi\chi}
N_\chi\over (N_\phi^2+N_\chi^2)} \Delta\chi \approx  0.14~. 
\label{statNG}
\end{equation}
Here we neglected the non-linear terms in the Sachs-Wolfe relation 
between ${\mathcal R}_{\rm c}$ and the observed temperature fluctuation
\begin{equation}
 {\delta T\over T} = -{{\mathcal R}_c\over 5}+O({\mathcal R}_c^2), 
\label{non-linearSW}
\end{equation} 
since the effect due to this non-linearity 
is suppressed by $N_\chi^2/N_{\chi\chi}$ and it can be 
absorbed by the redefinition of $N_{\chi\chi}$. 

On the other hand, this model predicts the local-type non-Gaussianity, 
\begin{equation}
 f_{\rm NL}={5\over 6}{N_{\chi\chi} N_{\chi}^2\over(N_\phi^2+N_\chi^2)^2}
=\frac{5}{24}\left({\Delta P_{\delta T}\over P_{\delta T}}\right)^2
\frac{1}{N_{\chi\chi}\Delta\chi^2}\approx
\frac{0.004}{N_{\chi\chi}\Delta\chi^2}
\left(\frac{\Delta P_{\delta T}/P_{\delta T}}{0.14}\right)^2
\,,
\label{fNL}
\end{equation}
where we have used Eq.~(\ref{statNG}) in the second and third equalities.
Furthermore, a quadrupolar anisotropy is induced by 
the second-order effect of $\Delta\chi$ which originates from 
the term $N_{\chi\chi}\Delta\chi^2$ in Eq.~(\ref{deltaN}).
Its magnitude is estimated as 
\begin{equation}
\left\vert a^{(\Delta\chi)}_{20}\right\vert
={2\sqrt{\pi}\over 15\sqrt{5}}|N_{\chi\chi}|
  \Delta\chi^2.
\label{ampa20}
\end{equation}
We note that this is not the so-called Grishchuk-Zel'dovich (GZ) 
effect~\cite{gz}. 
We also mention that although the functional dependence of the quadrupole 
is similar to that obtained in Ref.~\cite{Lyth}, our estimate gives a 
more robust bound while that in Ref.~\cite{Lyth} is based on a 
rather crude argument concerning the convergence of perturbative expansion
which cannot be directly compared with the observational data.
As mentioned earlier, if $\Delta{\mathcal R}_{\rm c}$ is composed of 
the super-curvature mode in the open inflation scenario, 
the constraint from the GZ effect can be evaded. 

Combining Eqs.~(\ref{fNL}) and (\ref{ampa20}), we find
\begin{eqnarray}
\left|a^{(\Delta\chi)}_{20}\right|\left|f_{\rm NL}\right|
\approx4.3\times10^{-4}
\left(\frac{\Delta P_{\delta T}/P_{\delta T}}{0.14}\right)^2.
\label{quadfnl}
\end{eqnarray}
Thus one cannot make both the quadrupole and local-type non-Gaussianity
very small simultaneously.
This gives a bound slightly tighter and more 
robust than that obtained in Eq.~(59) of Ref.~\cite{Lyth}.
Here we note that the contribution of $N_\phi^2$ does not 
relax the above constraint. 

From the observed quadrupole anisotropy, we may bound 
the induced quadrupole as
\begin{eqnarray}
 \left\vert a^{(\Delta\chi)}_{20}\right\vert\lesssim \beta
\sqrt{\langle a_{2m}^2\rangle}_{\rm obs}\equiv \beta \sqrt{C_2}~,
\label{ampbeta}
\end{eqnarray}
where the observed magnitude of the squared quadrupole temperature 
anisotropy is $C_2\approx 4.2\times 10^{-11}$ and 
we inserted a threshold number $\beta$ of order unity.
This implies the constraint 
\begin{equation}
{2\sqrt{\pi}\over 15\sqrt{5}}  
   \left\vert N_{\chi\chi}\right\vert
  \Delta\chi^2 \lesssim 
  \beta\sqrt{C_2} \,,
\label{ampQ}
\end{equation}
which should be satisfied in any kind of model.   
Inserting Eq.~(\ref{ampbeta}) into Eq.~(\ref{quadfnl}), we obtain 
\begin{equation}
  |f_{\rm NL}| \gtrsim 66\beta^{-1}\,. 
\end{equation}
The required value of $f_{\rm NL}$ is already difficult to reconcile with 
the Planck observation ($f_{\rm NL}=2.7\pm 5.8$)~\cite{Ade:2013ydc}. 
We emphasize that 
this constraint differs from the one discussed 
in the literature~\cite{Kamionkowski, Lyth}.

If we require the above constraint on the non-Gaussianity to be satisfied, 
we should take $\beta\gtrsim6$. 
This implies, from Eq.~(\ref{ampbeta}),
that $a^{(\Delta\chi)}_{20}$ must be accidentally cancelled by 
the completely independent CMB quadrupole contribution 
to give the observed small value of $C_2$.
Secondly, the value of $\Delta\chi$ must be fine-tuned so that 
$a^{(\Delta\chi)}_{20}$ is almost exactly $O(10^{-5})$. 
Thus it is manifest that any models of this kind require extreme
fine-tuning.

%added end=====================================================

\if0
Furthermore, when the 
temperature perturbation is dominated by 
the curvaton, there should be the 
dipole temperature perturbation $N_\chi\Delta\chi/5 \approx (1/6)(\Delta
P_{\delta T}/P_{\delta T}) f_{NL}^{-1}$ 
caused by $\Delta\chi$, which means that 
the peculiar velocity field relative to the CMB rest frame 
is shifted by the order of $\sim 3.5\times 10^{3} f_{NL}^{-1}$km/sec. 
This will be in conflict with the current observation of 
the magnitude and the direction of 
the CMB dipole, which is consistent with the clustering dipoles 
determined by the distribution of galaxies~\cite{Maller:2003en}. 
\fi

\section{A generic difficulty in models with two fields}

We found it difficult to construct a viable model to explain 
the dipolar statistical anisotropy in the context of 
simple curvaton models using the non-linear coupling $N_{\chi\chi}$.
One might wonder if using
the cross term $N_{\phi\chi}$ instead of $N_{\chi\chi}$ could
improve the situation. It turns out that this possibility  
for generating the dipolar statistical anisotropy is also almost ruled out.
This is because one can show that the amplitude of the dipole anisotropy 
$a_{10}^{(\Delta\chi)}$ inevitably becomes too large if the dipolar statistical 
anisotropy is generated by the cross term $N_{\phi\chi}$. 

In order to produce the dipolar statistical 
anisotropy, we must have 
$
 |({N_{\phi\chi}/N_\phi})\Delta\chi| \gtrsim 0.07.
$
However, within one expansion time during the inflation, $H^{-1}$,
the average value of the inflaton $\phi$ changes by $\sim \dot\phi H^{-1}$ 
, where a dot denotes the cosmic time derivative. 
Therefore, as long as the cross term $N_{\phi\chi}$ exists, 
this shift of the background value of $\phi$ generates $N_{\chi}$,
$$
\left|N_{\chi}\right|\gtrsim \left\vert N_{\phi\chi}{\dot\phi \over H}\right\vert
\approx\left\vert {N_{\phi\chi}\over N_\phi}\right\vert\,.
$$  
Hence the dipole temperature perturbation caused by $\Delta\chi$ 
becomes 
$$
\left\vert a^{(\Delta\chi)}_{10}\right\vert \approx {1\over 5}
\left|N_\chi\Delta\chi \right| \gtrsim {0.07\over 5}\approx 0.014~. 
%check
$$
This constraint is one order of magnitude larger than $1.2\times 10^{-3}$, 
which is obtained from the analysis 
of the Doppler boosting by Planck~\cite{Aghanim:2013suk}.  

%added begin=====================================================
We note that the above argument also applies to a single-field model, 
in which the role of the $\chi$-field is also played by the inflaton $\phi$. 
For this case, $N_{\phi\chi}$ and $\Delta\chi$ 
in the above argument are 
replaced with $N_{\phi\phi}$ and $\Delta\phi$, respectively.  
Consequently, single-field models are also in conflict with observation. 
%added end=====================================================

If we introduce the third field, it may be
possible to construct models that generate the dipolar statistical anisotropy 
using the non-linear terms in the $\delta N$ formula. 
However, as it seems difficult to avoid 
fine-tuning in such models, we do not pursue this 
direction further.

\section{The amplitude modulation of the fluctuations}
The above consideration teaches us that it is not easy 
to generate the dipolar statistical anisotropy by using cross terms 
in the expansion of the $\delta N$ formula, 
and simultaneously satisfy the observational constraint that 
the local-type non-Gaussianity is small. 
Therefore we pursue a different approach. 
We consider the possibility of modulating the amplitude of the quantum
fluctuations at horizon crossing without affecting the inflationary dynamics.

First we examine a model in which the Hubble expansion rate 
is modulated, but the number of $e$-folds $N$ is not affected by this
modulation.
A model Lagrangian is given by 
\begin{eqnarray}
L= -{1\over 2}(\nabla\phi)^2 - {1\over 2}
(\nabla\sigma)^2
  -V(\phi,\sigma)\,;\quad
V(\phi,\sigma)=W(\sigma)\,U(\phi)\,.  
\label{prodpot}
\end{eqnarray}
Here the $\sigma$-field is supposed to have a large amplitude of 
the super-curvature perturbation, but not to affect the inflationary dynamics. 
Namely, it is assumed that the curvature perturbation is generated by 
the inflaton $\phi$. This means $N=N(\phi)$.
On the other hand, the Hubble expansion rate $H$
is a function of not only $\phi$ but also $\sigma$, $H=H(\phi,\sigma)$. 
For $N=N(\phi)$, this implies $H=H(N,\sigma)$.
In other words, if we draw contours of $N=const.$ on the $(\phi,\sigma)$ plane, 
$H$ on the contour of a given value of $N$ may depend on $\sigma$.
Then the inflaton fluctuation at around $N=N_*$,
$
 \langle \delta\phi^2\rangle\approx {H^2(N_*,\sigma)}/{(2\pi)^2},
$
also depends on $\sigma$, 
where $N_*$ is
the value of the $e$-folding number at which the comoving scale of
the current Hubble horizon size left the horizon during inflation.

At first sight, for a potential of the form given 
by Eq.~(\ref{prodpot}), this model seems viable
because the $e$-folding number $N$ becomes $\sigma$-independent,
\begin{equation}
 N=\int_{\phi_{\rm f}}^\phi {{\rm d}\phi\over M_{\rm pl}^2} 
   {V(\phi,\sigma) \over V_\phi(\phi,\sigma)}
  =\int_{\phi_{\rm f}}^\phi {{\rm d}\phi\over M_{pl}^2} {U(\phi) 
      \over U_\phi(\phi)}\,,
\end{equation}
under the slow-roll approximation for $\phi$, 
where the subscript $\phi$ denotes
the $\phi$-derivative, $V_\phi=\partial_\phi V$, and
$\phi_{\rm f}$ is the value of $\phi$ at the end of inflation.

Unfortunately, however, the above formula for $N$ 
has non-negligible higher-order corrections. In fact, there always exists 
a correction of the form
\begin{eqnarray*}
   N_{\rm corr}\sim \frac{1}{6}\Bigl[\ln H^2\Bigr]^\phi_{\phi_{\rm f}}\,,
\end{eqnarray*}
which depends on $W(\sigma)$.
 Therefore the curvature 
perturbation induced by the super-curvature perturbation $\Delta\sigma$ 
becomes as large as $|N_\sigma\Delta\sigma|\approx
|(W_\sigma/W)\Delta\sigma/6| \approx 0.02$, where the
last equality follows from the fact that the magnitude of the 
dipolar anisotropy is given by $|(W_\sigma/W)\Delta\sigma/2|$. 
Hence this model produces
too large an amplitude of the dipole anisotropy $a_{10}$, which is already 
ruled out observationally~\cite{Ade:2013ydc}.

\section{A viable model}

Finally we propose a viable model. 
As before,
we assume the existence of a field $\sigma$ that
carries a super-curvature perturbation $\Delta\sigma$.
In addition, we introduce a curvaton $\chi$.
The idea is to modulate the dynamics of the curvaton 
$\chi$ by using the $\sigma$-field 
without affecting the inflationary dynamics.
This can be realized by assuming 
that the kinetic term of $\chi$ depends on $\sigma$. 
A model Lagrangian takes the form
\begin{equation}
L= -{1\over 2}(\nabla\phi)^2 -{1\over 2}m^2_\phi \phi^2
- {1\over 2}(\nabla\sigma)^2-{1\over 2}m^2_\sigma \sigma^2
-{1\over 2}f^2(\sigma)(\nabla\chi)^2  -{1\over 2}m^2_\chi \chi^2\,. 
\end{equation}
Owing to the function $f(\sigma)$, the amplitude of the fluctuations of $\chi$
is modulated as 
$\langle \delta\chi^2\rangle\approx H^2/\bigl(2\pi f(\sigma)\bigr)^2$.
With an appropriate choice of $f(\sigma)$, this model can easily explain
 the dipolar statistical anisotropy.
%added begin=====================================================

Assuming the hierarchy 
$
m^2_\sigma\gtrsim H^2\gg m^2_\phi\gg m^2_\chi
$
inside the bubble, one can make $\sigma$ decay quickly during inflation
so that it will not affect the dynamics during or after inflation
except for the modulation of the fluctuation amplitude $\delta\chi$ 
near or beyond the current Hubble horizon scale. 
We may also assume that the final curvature perturbation is 
dominated by the curvaton $\chi$
 and ignore the fluctuations of the inflaton $\phi$.
Then, we can easily suppress the local-type non-Gaussianity. 
%added end=======================================================
In this model the origin of the modulation is
assumed to die out during the inflation. Therefore, the amplitude of the 
dipolar statistical anisotropy becomes smaller on smaller length scales.
This seems to be favored by observational data~\cite{hirata,Flender:2013jja}.  

\ack
We would like to thank M. Bucher, J. Garriga, R. Gobbetti and J. Soda for
insightful discussions and comments.
This work was supported in part by 
the Grant-in-Aid for Scientific Research (Nos. 21244033, 21111006,
24103006 and 24103001), and in part by PHY-0855447 from the National Science 
Foundation. SK is grateful for the hospitality extended to her
by the Yukawa Institute for Theoretical Physics (YITP), Kyoto University, 
during her stay when this work was initiated. 
This work was completed during the molecule-type YITP workshop
``The CMB and theories of the primordial universe".

\end{document}